\documentclass[useAMS,usenatbib]{mn2e}
\usepackage{graphicx}
\usepackage{epsfig}

\title[Progenitors of SNe Ia with long delay times]{The progenitors of Type Ia supernovae with long delay times}
\author[B. Wang, X.-D. Li and Z.-W. Han]
{Bo Wang,$^{\rm 1,3}$\thanks{E-mail: wangbo@ynao.ac.cn} Xiang-Dong Li$^{\rm 2}$ and Zhan-Wen Han$^{\rm 1}$\\
$^1$ National Astronomical Observatories/Yunnan Observatory, the
Chinese Academy of Sciences, Kunming 650011, China\\
$^2$ Department of Astronomy, Nanjing University, Nanjing 210093, China\\
$^3$ Graduate University of Chinese Academy of Sciences, Beijing
100049, China}

\begin{document}
\pagerange{\pageref{firstpage}--\pageref{lastpage}} \pubyear{2009}
\maketitle

\begin{abstract}\label{abstract}\label{firstpage}
The nature of the progenitors of Type Ia supernovae (SNe Ia) is
still unclear. In this paper, by considering the effect of the
instability of accretion disk on the evolution of white dwarf (WD)
binaries, we performed binary evolution calculations for about 2400
close WD binaries, in which a carbon--oxygen WD accretes material
from a main-sequence star or a slightly evolved subgiant star (WD +
MS channel), or a red-giant star (WD + RG channel) to increase its
mass to the Chandrasekhar (Ch) mass limit. According to these
calculations, we mapped out the initial parameters for SNe Ia in the
orbital period--secondary mass ($\log P^{\rm i}-M^{\rm i}_2$) plane
for various WD masses for these two channels, respectively. We
confirm that WDs in the WD + MS channel with a mass as low as
$0.61\,M_\odot$ can accrete efficiently and reach the Ch limit,
while the lowest WD mass for the WD + RG channel is $1.0\,\rm
M_\odot$. We have implemented these results in a binary population
synthesis study to obtain the SN Ia birthrates and the evolution of
SN Ia birthrates with time for both a constant star formation rate
and a single starburst. We find that the Galactic SN Ia birthrate
from the WD + MS channel is $\sim$$1.8\times 10^{-3}\ {\rm yr}^{-1}$
according to our standard model, which is higher than previous
results. However, similar to previous studies, the birthrate from
the WD + RG channel is still low ($\sim$$3\times 10^{-5}\ {\rm
yr}^{-1}$). We also find that about one third of SNe Ia from the WD
+ MS channel and all SNe Ia from the WD + RG channel can contribute
to the old populations ($\ga$1\,Gyr) of SN Ia progenitors.

\end{abstract}


\section{INTRODUCTION}\label{1. INTRODUCTION}

Type Ia supernovae (SNe Ia) play an important role in astrophysics,
especially in cosmology. They appear to be good cosmological
distance indicators and have been applied successfully to the task
of determining cosmological parameters (e.g. $\Omega$ and $\Lambda$:
Riess et al. 1998; Perlmutter et al. 1999). They are also a key part
of our understanding of galactic chemical evolution owing to the
main contribution of iron to their host galaxies (e.g. Greggio \&
Renzini 1983; Matteucci \& Greggio 1986). Despite their importance,
several key issues related to the nature of their progenitors and
the physics of the explosion mechanisms are still not well
understood (Hillebrandt \& Niemeyer 2000; R\"{o}pke \& Hillebrandt
2005; Podsiadlowski et al. 2008; Wang et al. 2008a), and no SN Ia
progenitor system has been conclusively identified pre-explosion.

There is a broad agreement that SNe Ia are thermonuclear explosions
of carbon--oxygen white dwarfs (CO WDs) in binaries (see the review
of Nomoto, Iwamoto \& Kishimoto 1997). Over the last few decades,
two competing progenitor models of SNe Ia were discussed frequently,
i.e. the double-degenerate (DD) and single-degenerate (SD) models.
Of these two models, the SD model (Whelan \& Iben 1973; Nomoto,
Thielemann \& Yokoi 1984; Fedorova, Tutukov \& Yungelson 2004; Han
2008) is widely accepted at present. It is suggested that the DD
model, which involves the merger of two CO WDs (Iben \& Tutukov
1984; Webbink 1984; Han 1998), likely leads to an accretion-induced
collapse rather than a SN Ia (Nomoto \& Iben 1985; Saio \& Nomoto
1985; Timmes, Woosley \& Taam 1994). For the SD model, the companion
is probably a main-sequence (MS) star or a slightly evolved subgiant
star (WD + MS channel), or a red-giant star (WD + RG channel), or
even a He star (WD + He star channel) (Hachisu, Kato \& Nomoto 1996;
Li $\&$ van den Heuvel 1997; Hachisu et al. 1999a; Langer et al.
2000; Han $\&$ Podsiadlowski 2004, 2006; Wang et al. 2009a,b).
Although the SD model is currently a favorable progenitor model of
SNe Ia, any single channel of the model cannot account for the
birthrate inferred observationally. Note that, some recent
observations have indirectly suggested that at least some SNe Ia can
be produced by a variety of different progenitor systems (e.g.
Hansen 2003; Ruiz-Lapuente et al. 2004; Patat et al. 2007; Voss \&
Nelemans 2008; Wang et al. 2008b; Justham et al. 2009).

At present, various progenitor models of SNe Ia can be examined by
comparing the distribution of the delay time (between the star
formation and SN Ia explosion) expected from a progenitor channel
with that of observations (e.g. Chen $\&$ Li 2007; Meng, Chen \& Han
2009; L\"{u} et al. 2009; Ruiter, Belczynski \& Fryer 2009).
Mannucci, Della Valle \& Panagia (2006) argued for the existence of
two separate SN Ia populations, a `prompt' component with a delay
time less than $\sim$100\,Myr, and a `delayed' component with a
delay time $\sim$3\,Gyr. By investigating the star formation history
(SFH) of 257 SN Ia host galaxies, Aubourg et al. (2008) found
evidence of a short-lived population of SN Ia progenitors with
lifetimes less than 180\,Myr. Botticella et al. (2008) and Totani et
al. (2008) analyzed host galaxies of SNe Ia and concluded that a
substantial fraction of SNe Ia must have long delay times on the
order of 2$-$3\,Gyr. Moreover, Schawinski (2009) recently
constrained the minimum delay time of 21 nearby SNe Ia by
investigating their host galaxies (early-type galaxies). The study
showed that these early-type host galaxies lack `prompt' SNe Ia with
a delay time less than $\sim$100\,Myr and that $\sim$70 per cent SNe
Ia have minimum delay times of 275\,Myr$-$1.25\,Gyr, while at least
20 per cent SNe Ia have minimum delay times of at least 1\,Gyr at 95
per cent confidence and two of these four SNe Ia are likely older
than 2\,Gyr.

For SNe Ia with the short delay times, Wang et al. (2009a) studied a
WD + He star channel to produce SNe Ia, in which a CO WD accretes
material from a He MS star or a He subgiant to increase its mass to
the Chandrasekhar (Ch) mass. The study showed the parameter spaces
for the progenitors of SNe Ia. By using a detailed binary population
synthesis (BPS) approach, Wang et al. (2009b, WCMH09) found that the
Galactic SN Ia birthrate from this channel is $\sim$$0.3\times
10^{-3}\ {\rm yr}^{-1}$, and that this channel can produce SNe Ia
with short delay times ($\sim$45$-$140\,Myr). For SNe Ia with the
long delay times ($\ga$1\,Gyr), this requires that the mass of the
companion should be $\la$$2\,M_{\odot}$.

Recently, Xu \& Li (2009) emphasized that the mass-transfer through
the Roche lobe overflow (RLOF) in the evolution of WD binaries may
become unstable (at least during part of the mass-transfer
lifetime), i.e. the mass-transfer rate is not equivalent to the
mass-accretion rate onto the WD. This important feature has been
ignored in nearly all of the previous theoretical works on SN Ia
progenitors except for King, Rolfe \& Schenker (2003)
\footnote{King, Rolfe \& Schenker (2003) adopted a similar method in
Li \& Wang (1998) to produce SNe Ia with long period dwarf novae in
a semi-analytic approach.} and Xu \& Li (2009), who inferred that
the mass-accretion rate onto the WD during dwarf nova outbursts can
be sufficiently high to allow steady nuclear burning of the accreted
matter and growth of the WD mass. In particular, the study of Xu \&
Li (2009) enlarges the region of SN Ia parameter spaces. However,
they only give the SN Ia parameter spaces with WD initial masses of
0.8 and $1\,M_{\odot}$. More detailed work is obviously needed to
investigate the influence of the accretion disk on the final results
and to give SN Ia birthrates and delay times.

Including the effect of the instability of accretion disk on the
evolution of WD binaries, the purpose of this paper is to study the
WD + MS and WD + RG channels towards SNe Ia comprehensively and
systematically, and then to determine the parameter spaces for SNe
Ia, which can be used in BPS studies. In Section 2, we describe the
numerical code for the binary evolution calculations and the grid of
the binary models. The binary evolutionary results are shown in
Section 3. We describe the BPS method in Section 4 and present the
BPS results in section 5. Finally, a discussion is given in Section
6, and a summary in Section 7.

\section{BINARY EVOLUTION CALCULATIONS}\label{2. BINARY EVOLUTION CALCULATIONS}
In our WD binary models, the lobe-filling star is a MS star or a
subgiant star (WD + MS channel), or a red-giant star (WD + RG
channel). The star transfers some of its material onto the surface
of the WD, which increases the mass of the WD as a consequence. If
the WD grows to 1.378$\,M_{\odot}$, we assume that it explodes as a
SN Ia.

\subsection{Stellar evolution code}
We use Eggleton's stellar evolution code (Eggleton 1971, 1972, 1973)
to calculate the WD binary evolutions. The code has been updated
with the latest input physics over the last three decades (Han,
Podsiadlowski \& Eggleton 1994; Pols et al. 1995, 1998). RLOF is
treated within the code described by Han, Tout \& Eggleton (2000).
We set the ratio of mixing length to local pressure scale height,
$\alpha=l/H_{\rm p}$, to be 2.0 and set the convective overshooting
parameter, $\delta_{\rm OV}$, to be 0.12 (Pols et al. 1997;
Schr$\ddot{\rm o}$der, Pols \& Eggleton 1997), which roughly
corresponds to an overshooting length of $\sim$0.25 pressure
scaleheights ($H_{\rm P}$). The opacity tables are compiled by Chen
\& Tout (2007) from Iglesias \& Rogers (1996) and Alexander \&
Ferguson (1994). In our calculations we use a typical Pop I
composition with H abundance $X=0.70$, He abundance $Y=0.28$ and
metallicity $Z=0.02$.

\subsection{Mass-accretion rates}
During the mass-transfer through the RLOF, the accreting material
can form an accretion disk surrounding the WD, which may become
thermally unstable when the effective temperature in the disk falls
below the H ionization temperature $\sim$6500\,K (e.g. van Paradijs
1996; King et al. 1997; Lasota 2001). This also corresponds to a
critical mass-transfer rate below which the disk is unstable.
Similar to the work of Xu \& Li (2009), we also set the critical
mass-transfer rate for a stable accretion disk to be
\begin{equation}
\dot{M}_{\rm cr, disk} \simeq 4.3\times 10^{-9}\,(P_{\rm orb}/4\,\rm
hr)^{1.7}\,\rm M_{\odot}\,yr^{-1},
\end{equation}
for WD accretors (van Paradijs 1996), where $P_{\rm orb}$ is the
orbital period. The locations of various types of cataclysmic
variable stars (e.g. the UX UMa, U Gem, SU UMa, and Z Cam systems)
in a ($P_{\rm orb}$, $\dot{M}_{\rm cr, disk}$) diagram are well
described by this expression (Smak 1983; Osaki 1996; van Paradijs
1996).

If the mass-transfer rate, $|\dot{M}_2|$, is higher than the
critical value $\dot{M}_{\rm cr, disk}$, we assume that the
accretion disk is stable and the WD accretes smoothly at a rate
$\dot{M}_{\rm acc}=|\dot{M}_2|$; otherwise the WD accretes only
during outbursts and the mass-accretion rate is $\dot{M}_{\rm
acc}=|\dot{M}_2|/d$, where $d$ is the duty cycle. The mass-accretion
rate is $\dot{M}_{\rm acc}=0$ during quiescence. King, Rolfe \&
Schenker (2003) showed that for typical values of the duty cycle
$\sim$0.1 to a few $10^{-3}$ the accretion rates onto the WD during
dwarf nova outbursts can be sufficiently high to allow steady
nuclear burning of the accreted matter. The limits on the duty
cycles of dwarf nova outbursts come from observations (Warner 1995):
(1) The outburst intervals for each object are quasi-periodic, but
within the dwarf nova family, the intervals can range from days to
decades. (2) The lifetime of an outburst is typically from 2 to 20
days and is correlated with the outburst interval. The
quasi-periodicity of the dwarf nova outbursts allows to use a single
duty cycle to roughly describe the change in the mass-transfer rate,
though we note that this is a simplification of the real,
complicated processes. In this paper we set the duty cycle to be
0.01 (we will discuss the effects of varying the assumed value in
the discussion section).

\subsection{Mass-growth rates}
Instead of solving stellar structure equations of a WD, we adopt the
prescription of Hachisu et al. (1999a) for the mass-growth of a CO
WD by accretion of H-rich material from its companion. The
prescription is given below. If the mass-accretion rate of the WD,
$\dot{M}_{\rm acc}$, is above a critical rate, $\dot{M}_{\rm
cr,WD}$, we assume that the accreted H steadily burns on the surface
of the WD and that the H-rich material is converted into He at a
rate $\dot{M}_{\rm cr,WD}$. The unprocessed matter is assumed to be
lost from the system as an optically thick wind at a mass-loss rate
$\dot{M}_{\rm wind}=|\dot{M}_{\rm 2}|-\dot{M}_{\rm cr,WD}$. The
critical mass-accretion rate is
 \begin{equation}
 \dot{M}_{\rm cr,WD}=5.3\times 10^{\rm -7}\frac{(1.7-X)}{X}(M_{\rm
 WD}/M_{\odot}-0.4) \,M_{\odot}\,yr^{-1},
  \end{equation}
where $X$ is the H mass fraction and $M_{\rm
 WD}$ is the mass of the accreting WD.

The following assumptions are adopted when $|\dot{M}_{\rm acc}|$ is
smaller than $\dot{M}_{\rm cr,WD}$.

(1) When $|\dot{M}_{\rm acc}|$ is less than $\dot{M}_{\rm cr,WD}$
but higher than $\frac{1}{2}\dot{M}_{\rm cr,WD}$, the H-shell
burning is steady and no mass is lost from the system.

(2) When $|\dot{M}_{\rm acc}|$ is lower than
$\frac{1}{2}\dot{M}_{\rm cr,WD}$ but higher than
$\frac{1}{8}\dot{M}_{\rm cr,WD}$, a very weak H-shell flash is
triggered but no mass is lost from the system.

(3) When $|\dot{M}_{\rm acc}|$ is lower than
$\frac{1}{8}\dot{M}_{\rm cr,WD}$, the H-shell flash is so strong
that no material is accumulated onto the surface of the WD.

We define the mass-growth rate of the He layer under the H-shell
burning as
 \begin{equation}
 \dot{M}_{\rm He}=\eta _{\rm H}|\dot{M}_{\rm acc}|,
  \end{equation}
where $\eta _{\rm H}$ is the mass-accumulation efficiency for
H-shell burning. According to the assumptions above, the values of
$\eta _{\rm H}$ are:
 \begin{equation}
\eta _{\rm H}=\left\{
 \begin{array}{ll}
 \dot{M}_{\rm cr,WD}/|\dot{M}_{\rm acc}|, & |\dot{M}_{\rm acc}|> \dot{M}_{\rm
 cr,WD},\\
 1, & \dot{M}_{\rm cr,WD}\geq |\dot{M}_{\rm acc}|\geq\frac{1}{8}\dot{M}_{\rm
 cr,WD},\\
 0, & |\dot{M}_{\rm acc}|< \frac{1}{8}\dot{M}_{\rm cr,WD}.
\end{array}\right.
\end{equation}

When the mass of the He layer reaches a certain value, He is assumed
to be ignited. If He-shell flashes occur, a part of the envelope
mass is assumed to be blown off. The mass-growth rate of WDs in this
case is linearly interpolated from a grid computed by Kate \&
Hachisu (2004), where a wide range of WD mass and mass-accretion
rate were calculated in the He-shell flashes.

We define the mass-growth rate of the CO WD, $\dot{M}_{\rm CO}$, as
 \begin{equation}
 \dot{M}_{\rm CO}=\eta_{\rm He}\dot{M}_{\rm He}=\eta_{\rm He}\eta_{\rm
 H}|\dot{M}_{\rm acc}|,
  \end{equation}
where $\eta_{\rm He}$ is the mass-accumulation efficiency for
He-shell flashes.

\subsection{Orbital angular momentum losses}
The evolution of these WD binaries is driven by the nuclear
evolution of the donor stars, and the change of the orbital angular
momentum of the binaries is mainly caused by wind mass-loss from the
WD. We assume that the mass lost from these binaries carries away
the same specific orbital angular momentum of the WD (the mass-loss
in the donor's wind is supposed to be negligible, but its effect on
the change of the orbital angular momentum, i.e. magnetic braking,
is included, e.g. Li $\&$ van den Heuvel 1997).

For the magnetic braking (MB) effect, we adopt the description of
angular momentum loss from Sills, Pinsonneault \& Terndrup (2000),
\begin{equation}
\frac{{\rm d}J_{\rm{MB}}}{{\rm d}t}=\left\{
 \begin{array}{lc}
-K \omega^3 \left(\frac{R_2}{R_{\odot}}\right)^{0.5}
\left(\frac{M_2}{M_{\odot}}\right)^{-0.5},  \hspace{20pt} \omega \leq \omega_{\rm{crit}},\\
-K \omega_{\rm{crit}}^2 \omega
\left(\frac{R_2}{R_{\odot}}\right)^{0.5}
\left(\frac{M_2}{M_{\odot}}\right)^{-0.5},  \ \omega >
\omega_{\rm{crit}},\\
\end{array}\right.
\end{equation}
where $K=2.7\times10^{47}$ g\,cm$^2$\,s (Andronov, Pinsonneault \&
Sills 2003), $\omega$ is the angular velocity of the binary, and
$\omega_{\rm{crit}}$ is the critical angular velocity at which the
angular momentum loss rate reaches a saturated state (Krishnamurthi
et al. 1997). Following the suggestion of Podsiadlowski, Rappaport
\& Pfahl (2002), we also add an ad hoc factor ${\rm
exp}(-0.02/q_{\rm{conv}}+1) \; \rm{if}\;q_{\rm{conv}}<0.02$ in
Equation (6), where $q_{\rm{conv}}$ is the mass fraction of the
surface convective envelop, to reduce the MB effect when the
convective envelope becomes too small.

\subsection{Grid calculations}
We incorporate the prescriptions above into Eggleton's stellar
evolution code and follow the evolution of WD + MS systems. We have
calculated about 2400 WD + MS systems, and obtained a large, dense
model grid, in which the lobe-filling star is a MS star or a
subgiant star (WD + MS channel), or a RG star (WD + RG channel). The
initial mass of the donor star, $M_{\rm 2}^{\rm i}$, ranges from
0.5$-$4.0$\,M_{\odot}$; the initial mass of the CO WD, $M_{\rm
WD}^{\rm i}$, is from 0.61$-$1.20$\,M_{\odot}$; the initial orbital
period of the binary system, $P^{\rm i}$, changes from the minimum
value, at which a zero-age MS (ZAMS) star would fill its Roche lobe,
to $\sim$40\,d.

\section{BINARY EVOLUTION RESULTS} \label{3. BINARY EVOLUTION RESULTS}
\subsection{An example of binary evolution calculations}
\begin{figure*}
\centerline{\epsfig{file=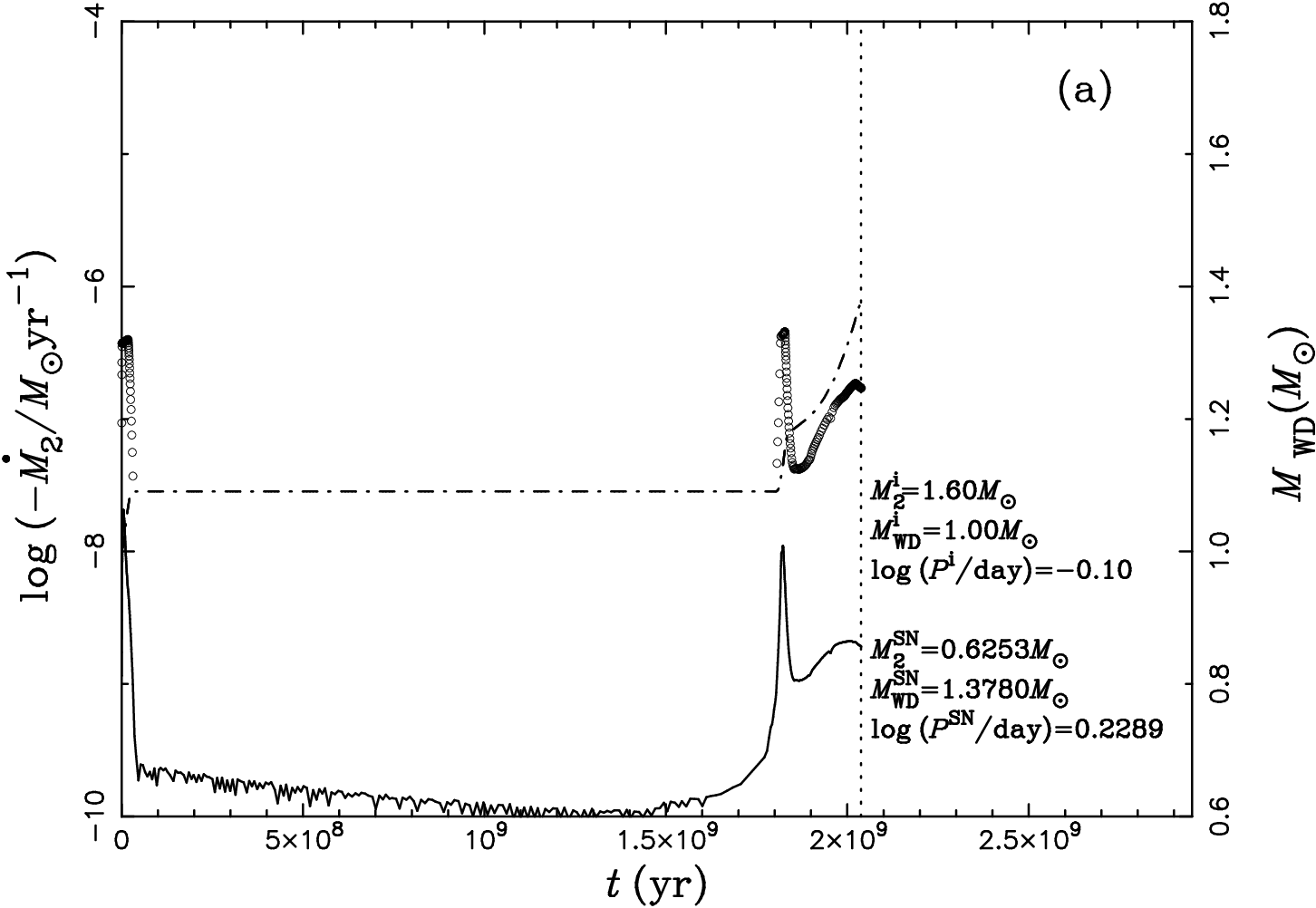,angle=270,width=8cm}\ \
\epsfig{file=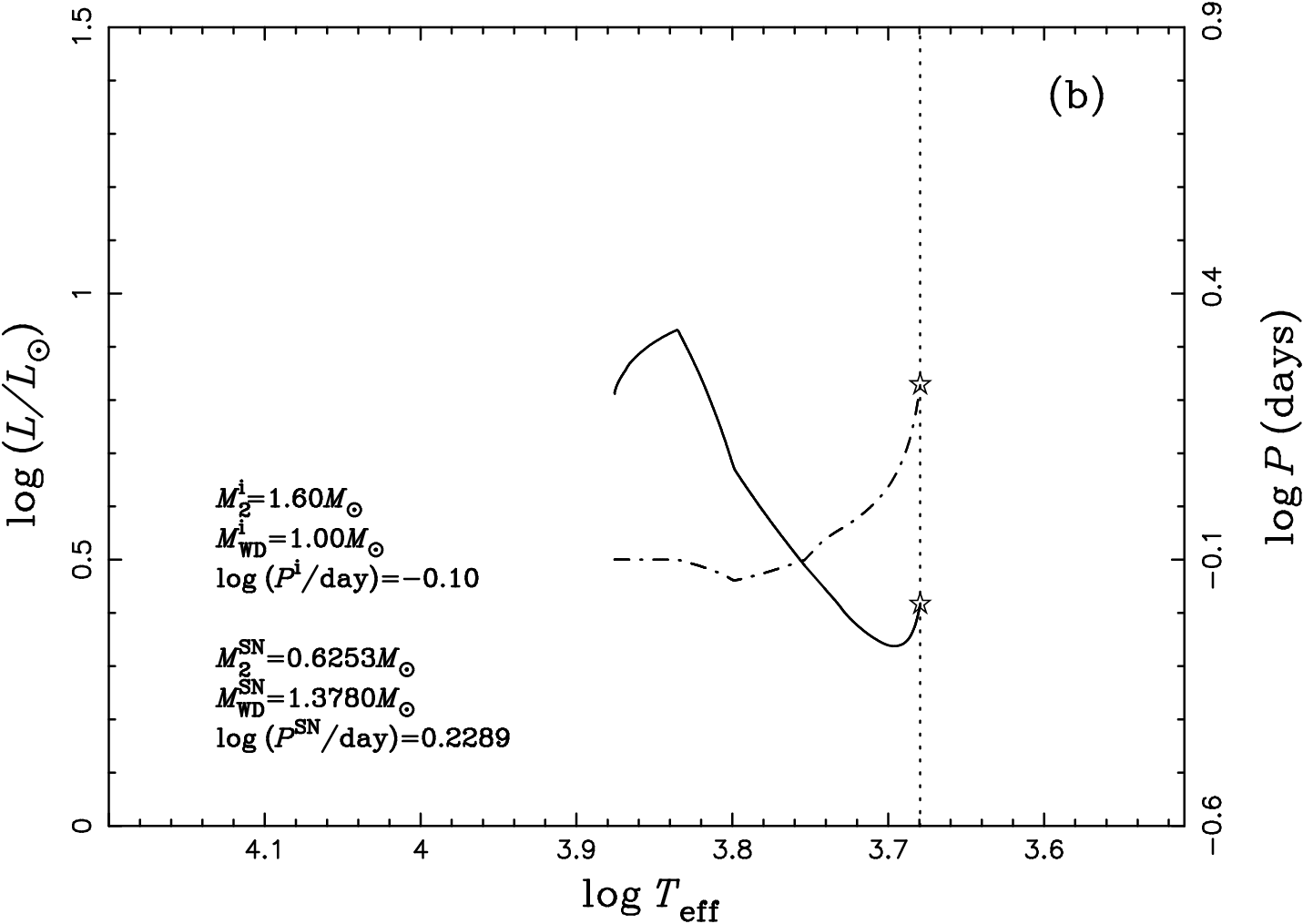,angle=270,width=8cm}} \caption{An example of
binary evolution calculations. In panel (a), the solid and
dash-dotted curves show $\dot M_2$ and $M_{\rm WD}$ varing with
time, respectively. The open circles represent the evolution of
$\dot M_{\rm CO}$ during outbursts. In panel (b), the evolutionary
track of the donor star is shown as a solid curve and the evolution
of orbital period is shown as a dash-dotted curve. Dotted vertical
lines in both panels and asterisks in panel (b) indicate the
position where the WD is expected to explode as a SN Ia. The initial
binary parameters and the parameters at the moment of the SN Ia
explosion are also given in these two panels.}
\end{figure*}

In Fig. 1, we present an example of binary evolution calculations.
Panel (a) shows the $\dot M_2$, $\dot M_{\rm CO}$ and $M_{\rm WD}$
varing with time, while panel (b) is the evolutionary track of the
donor star in the Hertzsprung-Russell diagram, where the evolution
of the orbital period is also shown. The binary shown in this case
is ($M_2^{\rm i}$, $M_{\rm WD}^{\rm i}$, $\log (P^{\rm i}/{\rm
day})$) $=$ (1.6, 1.0, $-$0.1), where $M_2^{\rm i}$, $M_{\rm
WD}^{\rm i}$ and $P^{\rm i}$ are the initial mass of the donor star
and the CO WD in solar masses, and the initial orbital period in
days, respectively. The donor star fills its Roche lobe on the MS
which results in case A mass-transfer. During the whole evolution,
the mass-transfer rate is lower than the critical value given by
Equation (1). Thus, the accretion disk experiences instability. The
mass-accretion rate $\dot M_{\rm acc}$ of the WD exceeds
$\frac{1}{8}\dot{M}_{\rm cr,WD}$ after the onset of RLOF, leading to
the mass-growth of the WD. With the continuous decreasing of
$|\dot{M}_{\rm 2}|$, when $\dot M_{\rm acc}$ drops below
$\frac{1}{8}\dot{M}_{\rm cr,WD}$ after about $5\times10^{7}$\,yr,
the H-shell flash is so strong that no material is accumulated onto
the surface of the WD. At about $1.7\times10^{9}$\,yr,
$|\dot{M}_{\rm 2}|$ increases again as the donor star evolves off
the MS and expands, allowing the mass-growth of the WD during
outbursts. When the WD grows to 1.378$\,M_{\odot}$, it explodes as a
SN Ia. At the SN explosion moment, the mass of the donor star is
$M^{\rm SN}_2=0.6253\,M_{\odot}$ and the orbital period $\log
(P^{\rm SN}/{\rm day})=0.2289$.

\subsection{Initial parameters for SN Ia progenitors}
\begin{figure*}
\epsfig{file=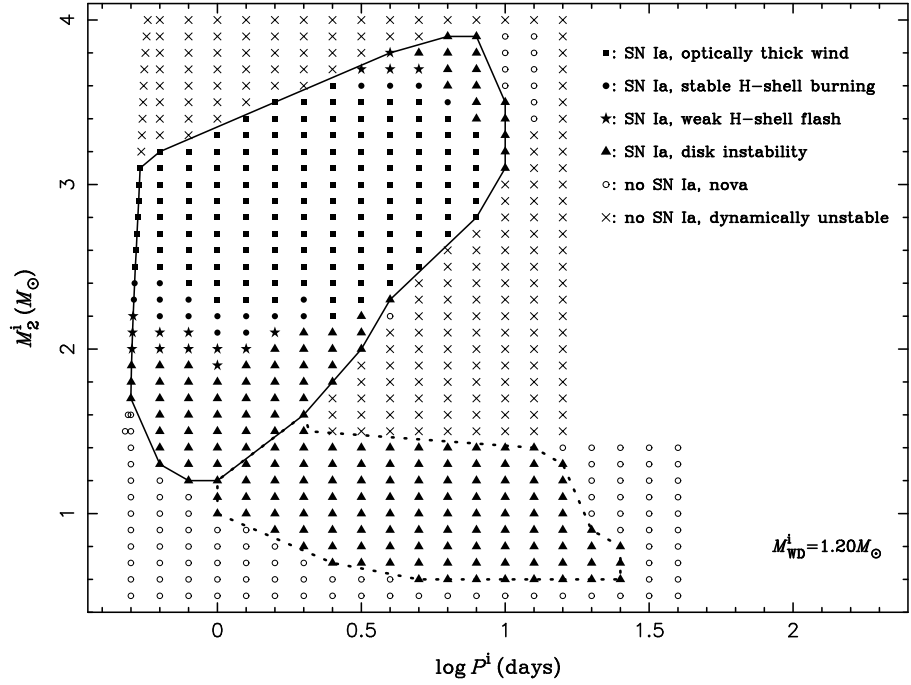,angle=270,width=12cm} \caption{Final outcomes of
the binary evolution calculations in the initial orbital
period--secondary mass ($\log P^{\rm i}$, $M^{\rm i}_2$) plane of
the CO WD + MS system for an initial WD mass of 1.2$\,M_{\odot}$.
Solid curves are for the contours from the WD + MS channel, while
dotted curves from the WD + RG channel. The filled squares, circles
and stars denote that the WD explodes as a SN Ia in the optically
thick wind phase, in the stable H-shell burning phase and in the
weak H-shell flash phase, respectively. The filled triangles
represent that the WD explodes as a SN Ia, in which the accretion
disk experiences instability. Open circles indicate systems that
experience novae, preventing the WD from reaching
1.378$\,M_{\odot}$, and crosses are those under dynamically unstable
mass-transfer.}
\end{figure*}

\begin{figure*}
\epsfig{file=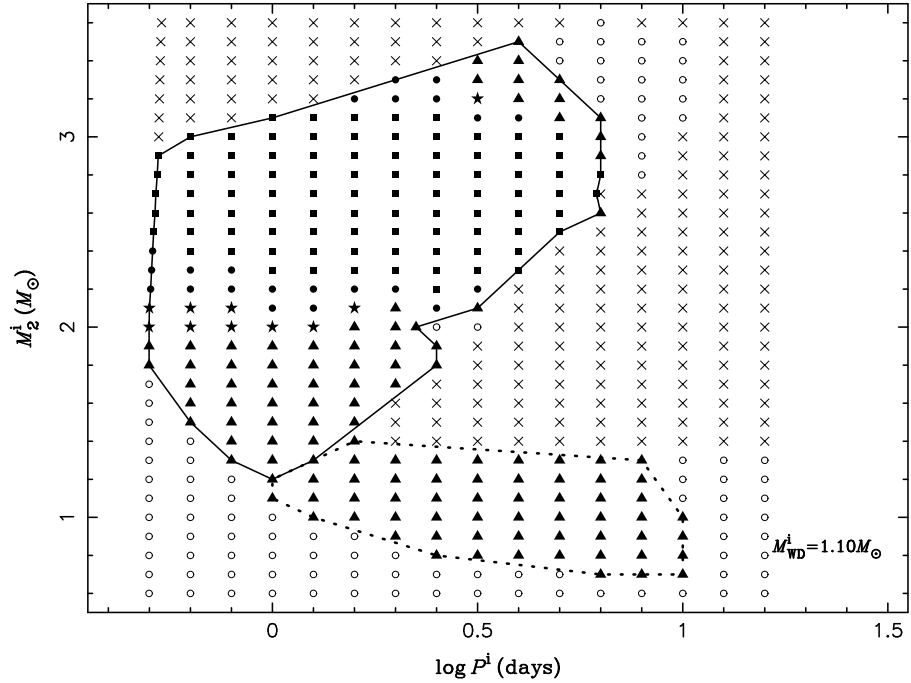,angle=270,width=12cm} \caption{Similar to Fig. 2,
but for an initial WD mass of 1.1$\,M_{\odot}$.}
\end{figure*}

\begin{figure*}
\centerline{\epsfig{file=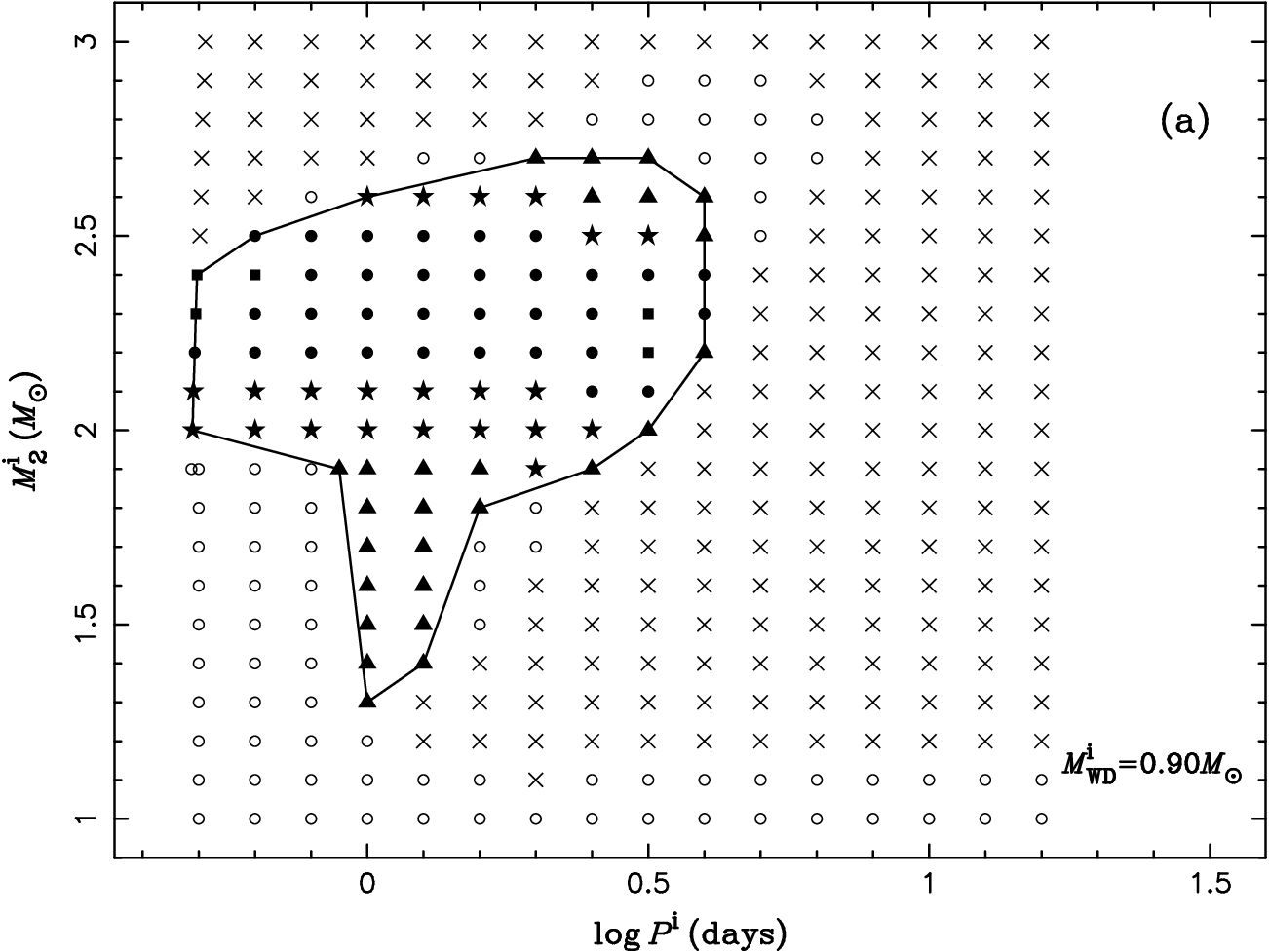,angle=270,width=8cm}\ \
\epsfig{file=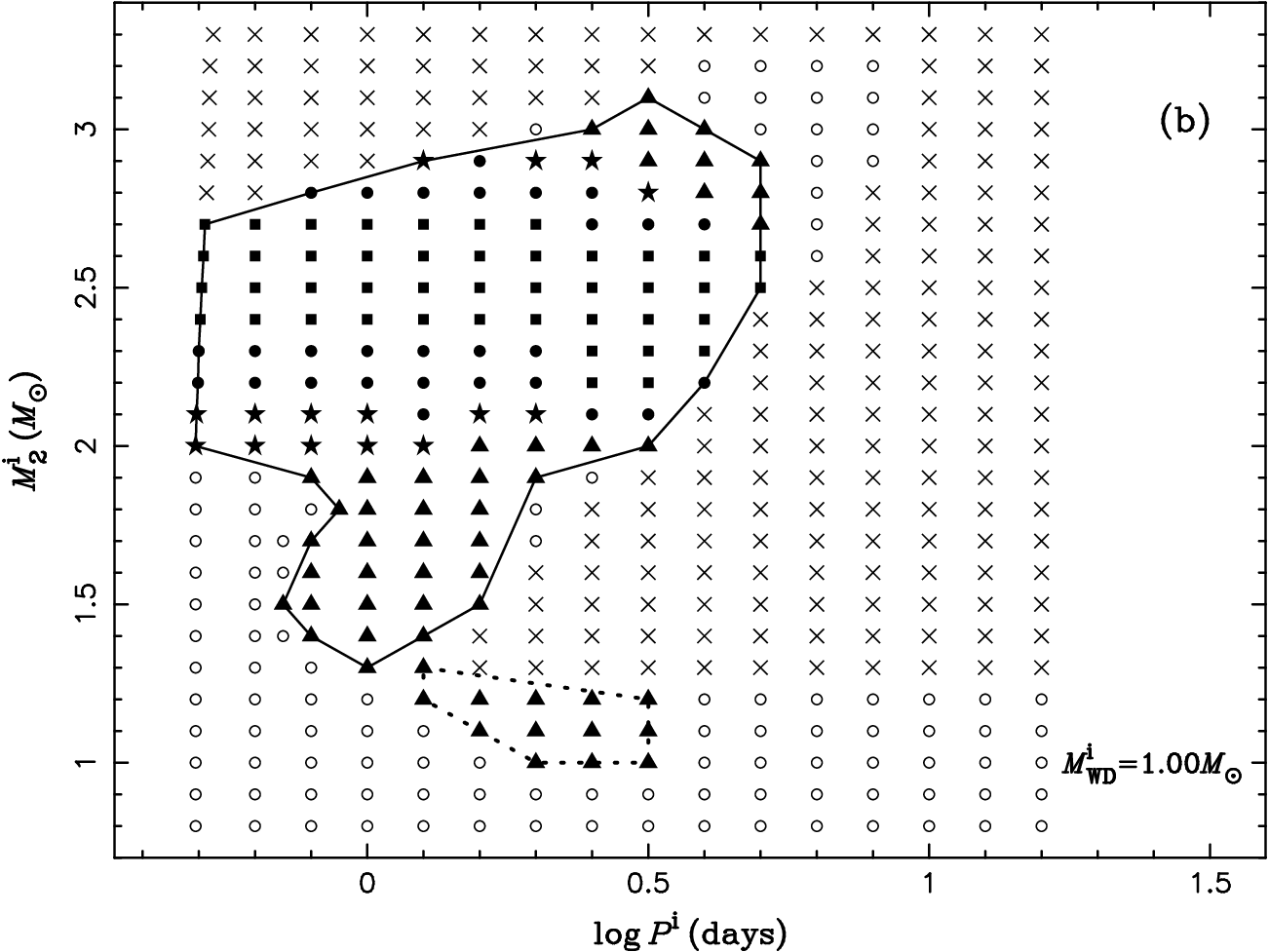,angle=270,width=8cm}} \caption{Similar to Fig.
2, but for initial WD masses of 0.9 and 1.0$\,M_{\odot}$.}
\end{figure*}

\begin{figure*}
\centerline{\epsfig{file=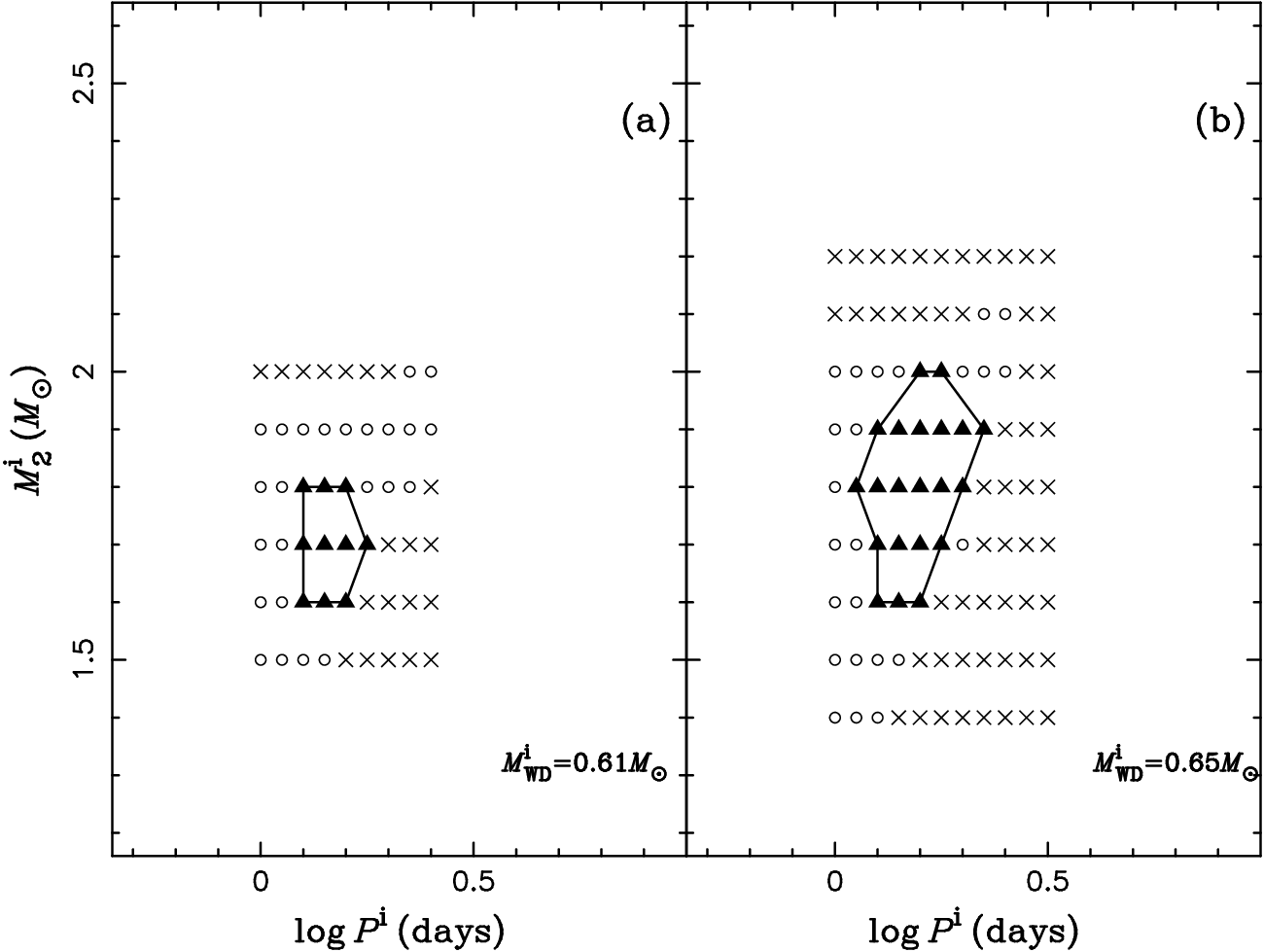,angle=270,width=8cm}\ \
\epsfig{file=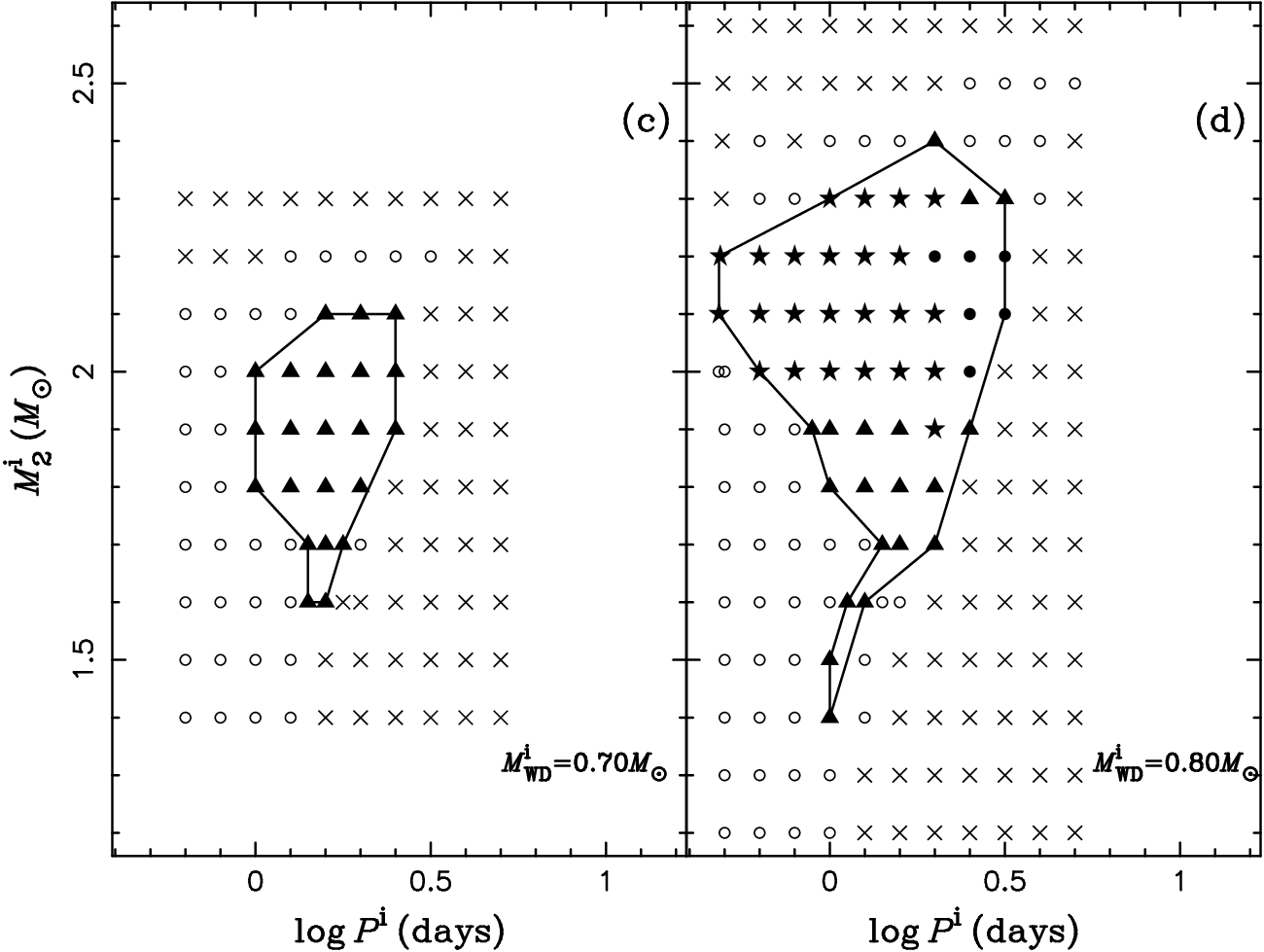,angle=270,width=8cm}} \caption{Similar to Fig.
2, but for initial WD masses of 0.61, 0.65, 0.7 and
0.8$\,M_{\odot}$.}
\end{figure*}

Figs 2-5 show the final outcomes of about 2400 binary evolution
calculations in the initial orbital period--secondary mass ($\log
P^{\rm i}-M^{\rm i}_2$) plane. The filled squares, circles and stars
denote that the WD explodes as a SN Ia in the optically thick wind
phase, in the stable H-shell burning phase and in the weak H-shell
flash phase, respectively. The filled triangles represent that the
WD explodes as a SN Ia, in which the accretion disk experiences
instability. Some binaries fail to produce SNe Ia owing to nova
explosions (which prevents the WD growing in mass; the open circles
in these figures) or dynamically unstable mass-transfer (resulting
in a common envelope; the crosses in these figures).

The contours of initial parameters for producing SNe Ia are also
presented in these figures (solid curves are for the contours from
the WD + MS channel, while dotted curves from the WD + RG channel).
The left boundaries of the contours in these figures (solid lines;
Figs 2-4 and panel (d) of Fig. 5) are set by the condition that RLOF
starts when the donor star is on the ZAMS, while systems beyond the
right boundary experience mass-transfer at a very high rate due to
the rapid expansion of the donor stars in the Hertzsprung gap (HG)
(solid lines) or RG stage (dotted lines) and they lose too much mass
via the optically thick wind, preventing the WDs increasing their
masses to the Ch mass. The upper boundaries are set mainly by a high
mass-transfer rate owing to a large mass-ratio. The lower boundaries
are constrained by the facts that the mass-transfer rate
$\dot{M}_{\rm 2}$ should be high enough to ensure that the WD can
grow in mass during outbursts and that the donor should be
sufficiently massive for enough mass to be transferred onto the WD.

In Fig. 6, we overlay the contours for SN Ia production in the
($\log P^{\rm i}$, $M^{\rm i}_2$) plane for different initial WD
masses. In panel (a), we give the contours of the WD + MS channel
for producing SNe Ia with various WD masses (i.e. $M_{\rm WD}^{\rm
i}=$ 0.61, 0.65, 0.7, 0.8, 0.9, 1.0, 1.1 and 1.2$\,M_{\odot}$). Note
that the enclosed region almost vanishes for $M_{\rm WD}^{\rm
i}=0.61\,M_{\odot}$, which is then assumed to be the minimum WD mass
for producing SNe Ia from this channel. As a comparison, we also
give the contours of the WD + RG channel in panel (b). If the
initial parameters of a WD binary are located in the contours, a SN
Ia is then assumed to be produced. Thus, these contours can be
expediently used in BPS studies. The data points and the
interpolation FORTRAN code for these contours can be supplied on
request by contacting BW.

\begin{figure*}
\centerline{\epsfig{file=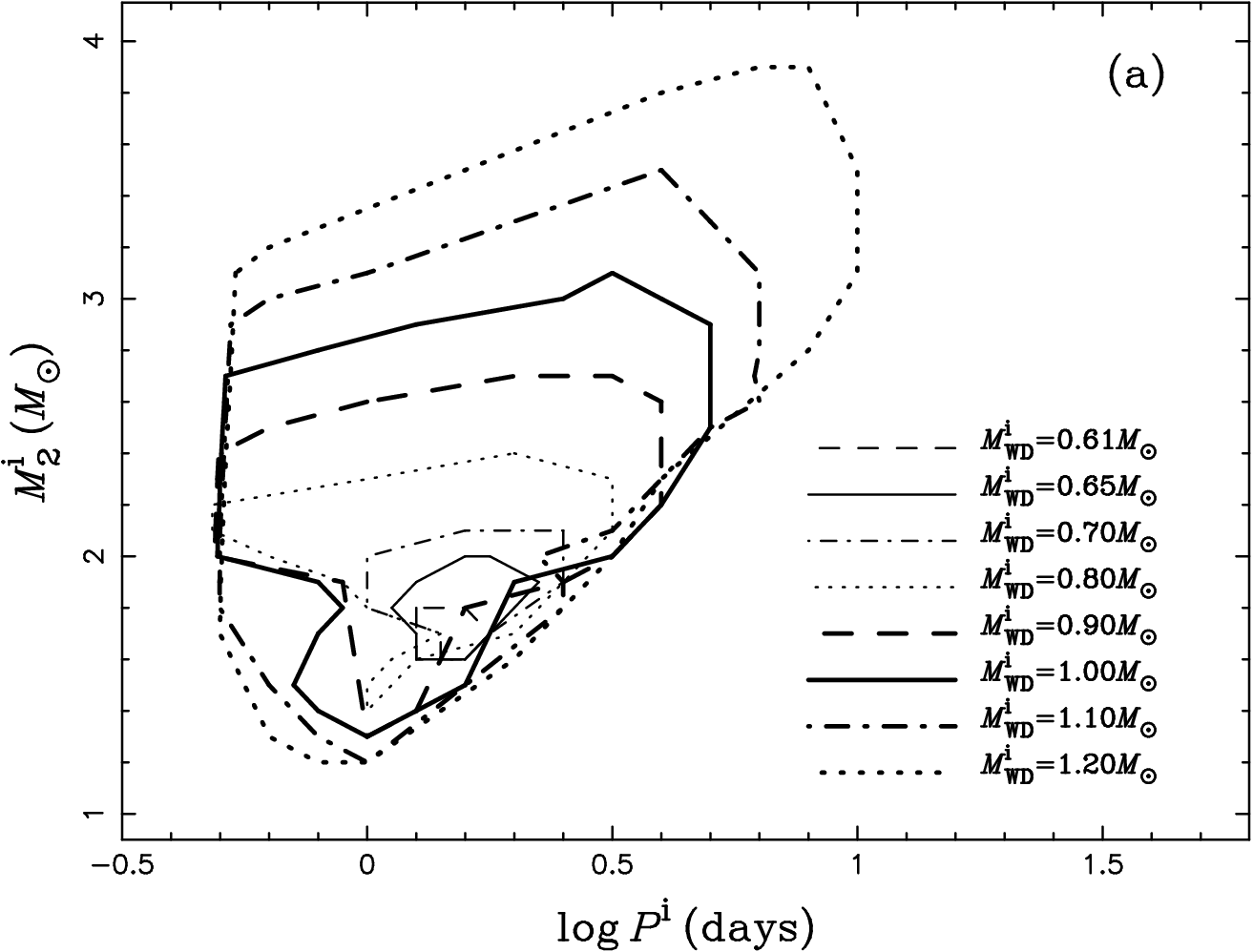,angle=270,width=8cm}\ \
\epsfig{file=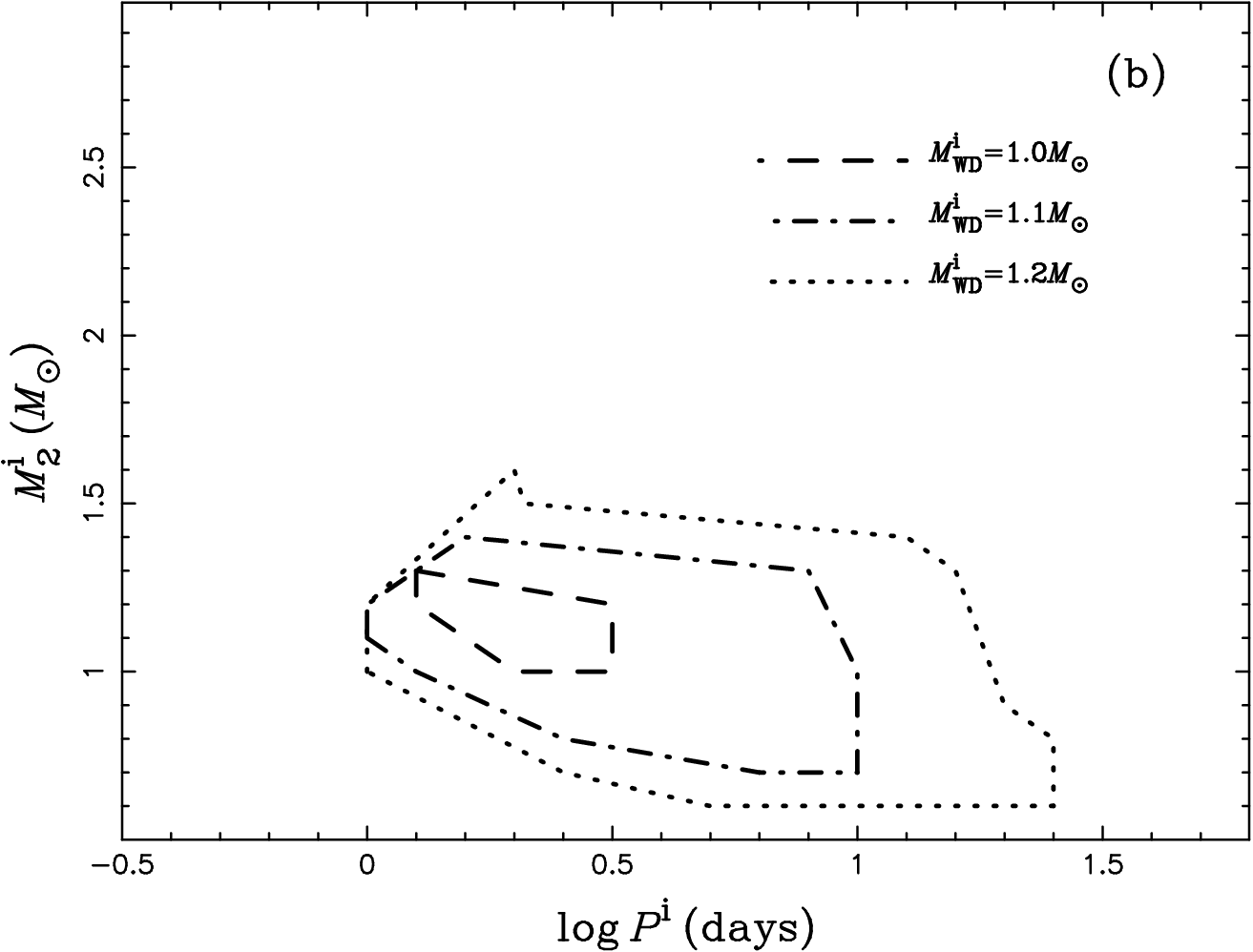,angle=270,width=8cm}} \caption{Panel (a):
regions in the initial orbital period--secondary mass plane ($\log
P^{\rm i}$, $M^{\rm i}_2$) for the WD + MS channel that produce SNe
Ia for initial WD masses of 0.61, 0.65, 0.7, 0.8, 0.9, 1.0, 1.1 and
1.2$\,M_{\odot}$. The region almost vanishes for $M_{\rm WD}^{\rm
i}=0.61\,M_{\odot}$. Panel (b): similar to the panel (a), but the
regions are for the WD + RG channel.}
\end{figure*}

\section{BINARY POPULATION SYNTHESIS} \label{4. BINARY POPULATION SYNTHESIS}
In order to investigate SN Ia birthrates and delay times for the WD
+ MS channel, we have performed a series of Monte Carlo simulations
in the BPS study. In each simulation, by using the Hurley's rapid
binary evolution code (Hurley, Pols \& Tout 2000, 2002), we have
followed the evolution of $1\times10^{\rm 7}$ sample binaries from
the star formation to the formation of the WD + MS systems according
to three evolutionary channels (Sect. 4.2). We assumed that, if the
parameters of a CO WD + MS system at the onset of the RLOF are
located in the SN Ia production regions (panel (a) of Fig. 6), a SN
Ia is produced. Note that, the method of the BPS study for the WD +
RG channel is similar to that of the WD + MS channel.

\subsection{Common envelope in binary evolution}
In the SD model of SNe Ia, the progenitor of a SN Ia is a close WD
binary system, which has most likely emerged from the common
envelope (CE) evolution (Paczy\'{n}ski 1976) of a giant binary
system. The CE ejection is still an open problem. Similar to the
work of WCMH09, we also use the standard energy equations (Webbink
1984) to calculate the output of the CE phase. The CE is ejected if
\begin{equation}
 \alpha_{\rm ce} \left( {G M_{\rm don}^{\rm f} M_{\rm acc} \over 2 a_{\rm f}}
- {G M_{\rm don}^{\rm i} M_{\rm acc} \over 2 a_{\rm i}} \right) = {G
M_{\rm don}^{\rm i} M_{\rm env} \over \lambda R_{\rm don}},
\end{equation}
where $\lambda$ is a structure parameter that depends on the
evolutionary stage of the donor, $M_{\rm don}$ is the mass of the
donor, $M_{\rm acc}$ is the mass of the accretor, $a$ is the orbital
separation, $M_{\rm env}$ is the mass of the donor's envelope,
$R_{\rm don}$ is the radius of the donor, and the indices ${\rm i}$
and ${\rm f}$ denote the initial and final values, respectively. The
right side of the equation represents the binding energy of the CE,
the left side shows the difference between the final and initial
orbital energy, and $\alpha_{\rm ce}$ is the CE ejection efficiency,
i.e. the fraction of the released orbital energy used to eject the
CE. For this prescription of the CE ejection, there are two highly
uncertain parameters (i.e. $\alpha_{\rm ce}$ and $\lambda$). As in
previous studies, we combine $\alpha_{\rm ce}$ and $\lambda$ into
one free parameter $\alpha_{\rm ce}\lambda$, and set it to be 0.5
and 1.5 (e.g. WCMH09).

\subsection{Evolutionary channels to WD + MS and WD + RG systems}
According to the evolutionary phase of the primordial primary at the
beginning of the first RLOF, there are three channels which can form
CO WD + MS systems and then produce SNe Ia.

(1) {\em He star channel.} The primordial primary first fills its
Roche lobe when it is in the HG (so-called early Case B
mass-transfer defined by Kippenhahn \& Weigert 1967). A CE may be
formed when the lobe-filling star evolves to the RG stage owing to a
large mass-ratio or a convective envelope of the mass donor star.
After the CE ejection, the primary becomes a He star and continues
to evolve. After the exhaustion of central He, the He star which now
contains a CO-core may fill its Roche lobe again due to expansion of
the He star itself, and transfer its remaining He-rich envelope to
the MS companion star, eventually leading to the formation of a CO
WD + MS system. For this channel, SN Ia explosions occur for the
ranges $M_{\rm 1,i}\sim4.0$$-$$7.0\,M_\odot$, $M_{\rm
2,i}\sim1.0$$-$$2.0\,M_\odot$, and $P^{\rm i} \sim 5$$-$30\,days,
where $M_{\rm 1,i}$, $M_{\rm 2,i}$ and $P^{\rm i}$ are the initial
mass of the primary and the secondary at ZAMS, and the initial
orbital period of a binary system.

(2) {\em EAGB channel.} If the primordial primary is on the early
asymptotic giant branch (EAGB, i.e. He is exhausted in the centre of
the star while thermal pulses have not yet started), a CE will be
formed because of dynamically unstable mass-transfer. After the CE
is ejected, the orbit decays and the primordial primary becomes a He
RG. The He RG may fill its Roche lobe and start mass-transfer, which
is likely stable and leaves a CO WD + MS system. For this channel,
SN Ia explosions occur for the ranges $M_{\rm
1,i}\sim2.5-6.5\,M_\odot$, $M_{\rm 2,i}\sim1.5-3.0\,M_\odot$ and
$P^{\rm i} \sim 200-900$\,days.

(3) {\em TPAGB channel.} The primordial primary fills its Roche lobe
at the thermal pulsing asymptotic giant branch (TPAGB) stage. A CE
is easily formed owing to dynamically unstable mass-transfer during
the RLOF. After the CE ejection, the primordial primary becomes a CO
WD, then a CO WD + MS system is produced. For this channel, SN Ia
explosions occur for the ranges $M_{\rm
1,i}\sim4.5$$-$$6.5\,M_\odot$, $M_{\rm
2,i}\sim1.5$$-$$3.5\,M_\odot$, and $P^{\rm i}\ga 1000$\,days.

Additionally, there is one channel which can form CO WD + RG systems
and then produce SNe Ia (i.e. the {\em TPAGB channel} above). After
a CO WD + MS system is produced via the {\em TPAGB channel}, the MS
companion star continues to evolve until the RG stage, i.e. a CO WD
+ RG system is formed. For the CO WD + RG systems, SN Ia explosions
occur for the ranges $M_{\rm 1,i}\sim5.0$$-$$6.5\,M_\odot$, $M_{\rm
2,i}\sim1.0$$-$$1.5\,M_\odot$, and $P^{\rm i}\ga 1500$\,days.

\subsection{Basic parameters for Monte Carlo simulations}

In the BPS study, the Monte Carlo simulation requires as input the
initial mass function (IMF) of the primary, the mass-ratio
distribution, the distribution of initial orbital separations, the
eccentricity distribution of binary orbit, and the star formation
rate (SFR) (e.g. Han et al. 2002, 2003; Han, Podsiadlowski \&
Lynas-Gray 2007; WCMH09).

(1) The IMF of Miller \& Scalo (1979) is adopted. The primordial
primary is generated according to the formula of Eggleton, Tout \&
Fitechett (1989),
\begin{equation}
M_{\rm 1}^{\rm p}=\frac{0.19X}{(1-X)^{\rm 0.75}+0.032(1-X)^{\rm
0.25}},
  \end{equation}
where $X$ is a random number uniformly distributed in the range [0,
1] and $M_{\rm 1}^{\rm p}$ is the mass of the primordial primary,
which ranges from 0.1\,$M_{\rm \odot}$ to 100\,$M_{\rm \odot}$. The
studies of the IMF by Kroupa, Tout \& Gilmore (1993) and Zoccali et
al. (2000) support this IMF.

(2) The initial mass-ratio distribution of the binaries, $q'$, is
quite uncertain for binary evolution. For simplicity, we take a
constant mass-ratio distribution (Mazeh et al. 1992; Goldberg \&
Mazeh 1994),
\begin{equation}
n(q')=1, \hspace{2.cm} 0<q'\leq1,
\end{equation}
where $q'=M_{\rm 2}^{\rm p}/M_{\rm 1}^{\rm p}$. This constant
mass-ratio distribution is supported by the study of Shatsky \&
Tokovinin (2002). As an alternative mass-ratio distribution we also
consider uncorrelated binary components, i.e. both binary components
are chosen randomly and independently from the same IMF
(uncorrelated).

(3) We assume that all stars are members of binaries and that the
distribution of separations is constant in $\log a$ for wide
binaries, where $a$ is separation and falls off smoothly at small
separation
\begin{equation}
a\cdot n(a)=\left\{
 \begin{array}{lc}
 \alpha_{\rm sep}(a/a_{\rm 0})^{\rm m}, & a\leq a_{\rm 0},\\
\alpha_{\rm sep}, & a_{\rm 0}<a<a_{\rm 1},\\
\end{array}\right.
\end{equation}
where $\alpha_{\rm sep}\approx0.07$, $a_{\rm 0}=10\,R_{\odot}$,
$a_{\rm 1}=5.75\times 10^{\rm 6}\,R_{\odot}=0.13\,{\rm pc}$ and
$m\approx1.2$. This distribution implies that the numbers of wide
binaries per logarithmic interval are equal, and that about 50\,per
cent of stellar systems have orbital periods less than 100\,yr (Han,
Podsiadlowski \& Eggleton 1995).

(4) A circular orbit is assumed for all binaries. The orbits of
semidetached binaries are generally circularized by the tidal force
on a timescale which is much smaller than the nuclear timescale.
Also, a binary is expected to become circularized during the RLOF.

(5) We simply assume a constant SFR over the last 15\,Gyr or,
alternatively, as a delta function, i.e. a single starburst. In the
case of the constant SFR, we calibrate the SFR by assuming that one
binary with a primary more massive than $0.8\,M_{\odot}$ is formed
annually (see Iben \& Tutukov 1984; Han, Podsiadlowski \& Eggleton
1995; Hurley, Pols \& Tout 2002). From this calibration, we can get
${\rm SFR}=5\,M_{\rm \odot}{\rm yr}^{-1}$ (e.g. Willems \& Kolb
2004), which successfully reproduces the $^{26}$Al 1.809-MeV
gamma-ray line and the core-collapse SN rate in the Galaxy (Timmes,
Diehl \& Hartmann 1997). It is obvious that for a galaxy with a
different SFR the SN Ia rates will be scaled. For the case of the
single starburst, we assume a burst producing $10^{11}\,M_{\odot}$
in stars, which is a typical mass of a galaxy. In fact, a galaxy may
have a complicated star formation history. We only choose these two
extremes for simplicity. A constant SFR is similar to the situation
of spiral galaxies (Yungelson \& Livio 1998; Han $\&$ Podsiadlowski
2004), while a delta function to that of elliptical galaxies or
globular clusters.

\section{The results of binary population synthesis} \label{5. The results of binary population synthesis}

\begin{table}
 \begin{minipage}{85mm}
 \caption{Galactic birthrates of SNe Ia for different simulation sets, where sets 1 and 4 are our standard models for the WD + MS and WD + RG channels, respectively.}
\begin{tabular}{cccccc}
\hline
Set & ${\rm Channel}$ & $\alpha_{\rm ce}\lambda$ & $n(q')$ & $\nu$ ($10^{-3}$\,yr$^{-1}$)\\
\hline
$1$ & ${\rm WD+MS}$ & $0.5$ & ${\rm Constant}$          & $1.793$\\
$2$ & ${\rm WD+MS}$ & $1.5$ & ${\rm Constant}$          & $1.415$\\
$3$ & ${\rm WD+MS}$ & $0.5$ & ${\rm Uncorrelated}$      & $0.275$\\
$4$ & ${\rm WD+RG}$ & $0.5$ & ${\rm Constant}$          & $0.028$\\
$5$ & ${\rm WD+RG}$ & $1.5$ & ${\rm Constant}$          & $0.026$\\
\hline
\end{tabular}
\end{minipage}
\end{table}

\begin{figure}
\includegraphics[width=6.cm,angle=270]{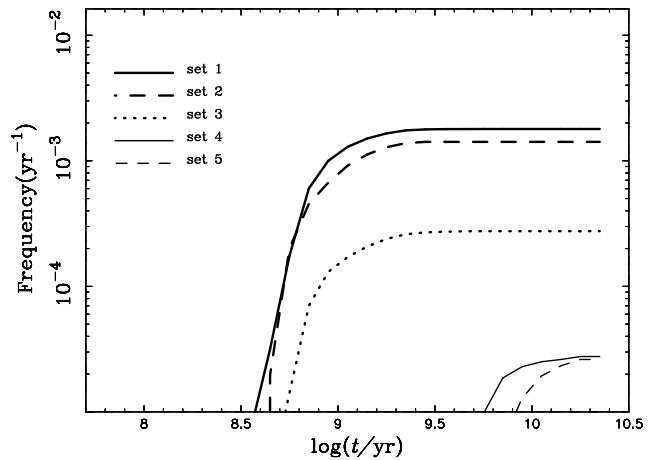}
 \caption{Evolution of Galactic birthrates of SNe Ia for a constant star
formation rate ($Z=0.02$, ${\rm SFR}=5\,M_{\rm \odot}{\rm
yr}^{-1}$). The key to the line-styles representing different sets
is given in the upper left corner.}
\end{figure}

We performed five sets of simulations (see Table 1) with metallicity
$Z=0.02$ to systematically investigate Galactic birthrates of SNe
Ia, where sets 1 and 4 are our standard models for the WD + MS and
WD + RG channels, respectively, and with the best choice of model
parameters (e.g. WCMH09). We vary the model parameters in the other
sets to examine their influences on the final results.

In Fig. 7, we show the evolution of Galactic SN Ia birthrate by
adopting $Z=0.02$ and ${\rm SFR}=5\,M_{\rm \odot}{\rm yr}^{-1}$ both
for the WD + MS channel (thick curves) and the WD + RG channel (thin
curves). The simulation for the WD + MS channel gives Galactic SN Ia
birthrate of $\sim$$1.8\times 10^{-3}\ {\rm yr}^{-1}$ according to
our standard model (set 1). The result is higher by a factor of
2$-$3 than previous studies by Han \& Podsiadlowski (2004, HP04)
($\sim$0.6$-$$0.8\times 10^{-3}\ {\rm yr}^{-1}$). This is mainly
attributed to the effect of the accretion disk instability, which
can increase the accretion rate onto the WD during outbursts,
leading to higher effective WD growth rates and a larger fraction of
systems becoming SN Ia progenitors. However, the result from this
work is still lower than that inferred observationally (i.e.
3$-$$4\times 10^{-3}\ {\rm yr}^{-1}$: van den Bergh \& Tammann 1991;
Cappellaro \& Turatto 1997). The simulation for a constant
mass-ratio distribution with $\alpha_{\rm ce}\lambda=1.5$ (set 2)
gives SN Ia birthrate of $\sim$$1.4\times 10^{-3}\ {\rm yr}^{-1}$,
which is lower than the case of $\alpha_{\rm ce}\lambda=0.5$ (the
binaries resulted from the CE ejections tend to have slightly closer
orbits for $\alpha_{\rm ce}\lambda=0.5$ and are more likely to
locate in the SN Ia production region). If we adopt a mass-ratio
distribution with uncorrelated binary components and $\alpha_{\rm
ce}\lambda=0.5$ (set 3), the SN Ia birthrate from the WD + MS
channel will be lower by an order of magnitude (see dotted curve in
Fig.7). This is because most of the donors in the WD + MS channel
are not very massive which has the consequence that WDs cannot
accrete enough mass to reach the Ch mass. Similar to previous
studies (e.g. Yungelson \& Livio 1998; HP04), the Galactic SN Ia
birthrate from the WD + RG channel is still low ($\sim$$3\times
10^{-5}\ {\rm yr}^{-1}$). The studies in this paper imply that the
WD + MS and WD + RG channels are only subclasses of SN Ia
production, and there may be some other channels or mechanisms also
contributing to SNe Ia, e.g. WD + He star channel or
double-degenerate channel. Especially, as mentioned by WCMH09, the
WD + He star channel can give a Galactic birthrate of
$\sim$$0.3\times 10^{-3}\ {\rm yr}^{-1}$, and is considered to be an
important channel to produce SNe Ia with short delay times. The SN
Ia birthrates shown in this figure seem to be so completely flat
after the first rise, for the specific reason see Sect. 3 of WCMH09.

\begin{figure}
\includegraphics[width=6.cm,angle=270]{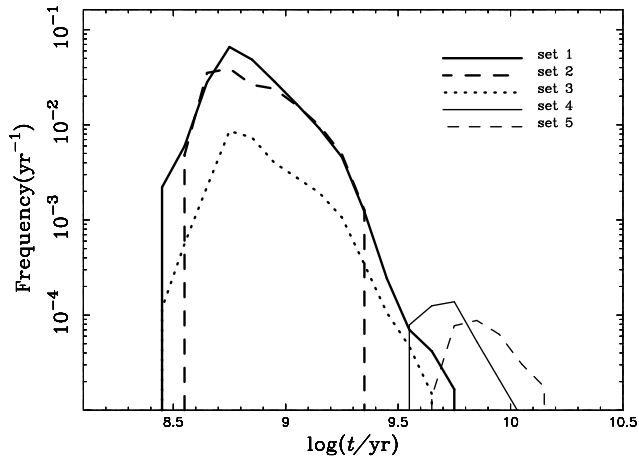}
 \caption{Similar to Fig. 7,
but for a single starburst with a total mass of $10^{\rm
11}M_{\odot}$.}
\end{figure}

Fig. 8 displays the evolution of SN Ia birthrates for a single
starburst with a total mass of $10^{11}\,M_{\odot}$. In the figure
we see that SNe Ia from the WD + MS channel (thick curves) have the
delay times of $\sim$250\,Myr$-$6.3\,Gyr (while SNe Ia from the WD +
RG channel have the delay times of $\ga$3\,Gyr; thin curves). This
figure also shows that a high value of $\alpha_{\rm ce}\lambda$
leads to a systematically later explosion time for these two
channels (e.g. the cases of $\alpha_{\rm ce}\lambda=0.5$ and 1.5 for
a constant mass-ratio distribution). This is because a high value of
$\alpha_{\rm ce}\lambda$ leads to wider WD binaries, and, as a
consequence, it takes a longer time for the companion to evolve to
fill its Roche lobe.

\begin{figure}
\includegraphics[width=6.cm,angle=270]{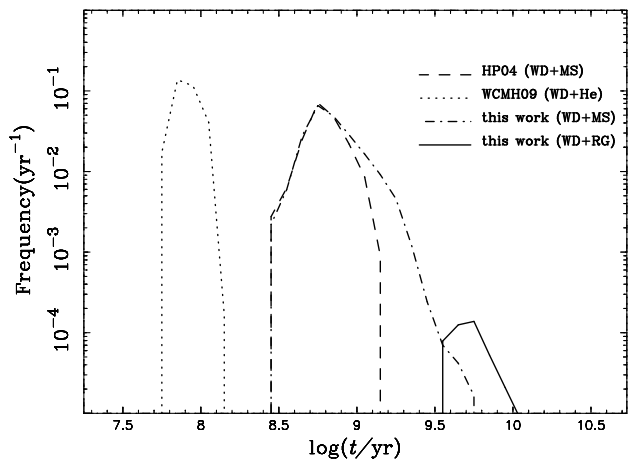}
 \caption{Same as Fig. 8,
but for different SD models. The SD models are from HP04 (dashed
curve), WCMH09 (dotted curve) and this work (dot-dashed curve is for
the WD + MS channel, while solid curve is for the WD + RG channel).
All curves are for a constant mass-ratio distribution and with
$\alpha_{\rm ce}\lambda=0.5$.}
\end{figure}

In Fig. 9, we compare the theoretical delay time distributions
(DTDs) of SNe Ia for different SD models. The SD models are from
HP04 (dashed curve), WCMH09 (dotted curve) and this work (dot-dashed
curve is for the WD + MS channel, while solid curve is for the WD +
RG channel). In this figure, the WD + MS channel of HP04 may
contribute to the SNe Ia with intermediate delay times
(100\,Myr$-$1\,Gyr: for a detailed discussion see Schawinski 2009).
The results of the WD + He star channel from WCMH09 may provide a
possible ways to explain the SNe Ia with short delay times
($\la$100\,Myr). In addition, for SNe Ia with the long delay times
($\ga$1\,Gyr), about one third of SNe Ia from the WD + MS channel
(dot-dashed curve) and all SNe Ia from the WD + RG channel (solid
curve) can contribute to the old populations of SN Ia progenitors.
Note, Chen \& Li (2007) studied the WD + MS channel by considering a
circumbinary disk which extracts the orbital angular momentum from
the binary through tidal torques. The study can also provide a
possible ways of producing such old SNe Ia ($\sim$1$-$3\,Gyr).
Finally, we emphasize that SNe Ia may be from several different
progenitor systems.

\section{DISCUSSION} \label{6. DISCUSSION}
The regions (Fig. 6 in this paper) for producing SNe Ia depend on
many uncertain input parameters, in particular for the duty cycle
which is poorly known. The main uncertainties lie in the facts that
it varies from one binary system to another and may evolve with the
orbital periods and mass-transfer rates (e.g. Lasota 2001; Xu \& Li
2009). This is the reason why we choose an intermediate value (0.01)
rather than other extreme ones (e.g. 0.1 or $10^{-3}$). However, we
also did some tests for a higher or lower value of the duty cycle.
We find that the variation of the duty cycle value will influence
the regions for producing SNe Ia, especially for the WD + RG channel
(for a high value (0.1), the regions will be shifted to higher
period; for a low value ($10^{-3}$), the regions will be shifted to
lower period). For these two extreme duty cycle values, the SN Ia
birthrate from the WD + RG channel will be lower due to the smaller
regions for producing SNe Ia.

In our binary evolution calculations we have not considered the
influence of rotation on the H-accreting WDs. The calculations of
Yoon, Langer \& Scheithauer (2004) showed that the He-shell burning
is much less violent when rotation is taken into account. This may
significantly increase the He-accretion efficiency ($\eta _{\rm He}$
in this paper). Meanwhile, the maximum stable mass of a rotating WD
may be above the standard Ch mass (i.e. the super-Ch mass model:
Uenishi, Nomoto \& Hachisu 2003; Yoon \& Langer 2005; Chen \& Li
2009). However, we mainly focus on the standard Ch mass explosions
of the accreting WDs in this work.

In our BPS studies we assume that all stars are in binaries and
about 50\,per cent of stellar systems have orbital periods less than
100\,yr. In fact, this is known to be a simplification, and the
binary fractions may depend on metallicity, environment, spectral
type, etc. If we adopt 40\,per cent of stellar systems have orbital
periods below 100\,yr by adjusting the parameters in Equation (10),
we estimate that the Galactic SN Ia birthrate from the WD + MS
channel will decrease to be $\sim$$1.4\times 10^{-3}\ {\rm yr}^{-1}$
according to our standard model (the birthrate from the WD + RG
channel will decrease to be $\sim$$2\times 10^{-5}\ {\rm yr}^{-1}$).
In addition, Umeda et al. (1999) concluded that the upper limit mass
of CO cores born in binaries is about 1.07$\,M_\odot$. If this value
is adopted as the upper limit of the CO WD, the SN Ia birthrate from
the WD + MS channel will decrease to be
$\sim$$1.7\times10^{-3}\,{\rm yr}^{-1}$ (the birthrate from the WD +
RG channel will decrease to be $\sim$$1\times 10^{-5}\ {\rm
yr}^{-1}$).

The Galactic SN Ia birthrate from the WD + RG channel is
$\sim$3$\times 10^{-5}\ {\rm yr}^{-1}$ according to our standard
model, which is low compared with observations, i.e. SNe Ia from
this channel may be rare. However, further study on this channel is
necessary, since this channel may explain some SNe Ia with long
delay times. In addition, it is suggested that, RS Oph and T CrB,
both recurrent novae are probable SN Ia progenitors and belong to
the WD + RG channel (e.g. Belczy$\acute{\rm n}$ski \& Mikolajewska
1998; Hachisu, Kato \& Nomoto 1999b; Sokoloski et al. 2006; Hachisu,
Kato \& Luna 2007). Meanwhile, by detecting Na I absorption lines
with low expansion velocities, Patat et al. (2007) suggested that
the companion of the progenitor of SN 2006X may be an early RG star.
Additionally, Voss \& Nelemans (2008) studied the pre-explosion
archival X-ray images at the position of the recent SN 2007on, and
they considered that its progenitor may be a WD + RG system.

\section{SUMMARY}\label{7. SUMMARY}

Employing Eggleton's stellar evolution code with the prescription of
Hachisu et al. (1999a) for the mass-accretion of CO WDs, and
including the effect of the instability of accretion disk on the
evolution of WD binaries, we performed binary evolution calculations
for about 2400 close WD binaries. The calculated results further
confirm that the disk instability could substantially increase the
mass-accumulation efficiency for accreting WDs, and cause the
possible SNe Ia to occur in systems with $\la$$2\,M_{\odot}$ donor
stars (see also King, Rolfe \& Schenker 2003; Xu \& Li 2009). We
find that the Galactic SN Ia birthrate from the WD + MS channel is
$\sim$$1.8\times 10^{-3}\ {\rm yr}^{-1}$ according to our standard
model, which is higher than previous results. However, similar to
previous studies, the birthrate from the WD + RG channel is still
low ($\sim$$3\times 10^{-5}\ {\rm yr}^{-1}$). We also find that
about one third of SNe Ia from the WD + MS channel and all SNe Ia
from the WD + RG channel can contribute to the old populations
($\ga$1\,Gyr) of SN Ia progenitors. The companion stars of SNe Ia
with long delay times in this work would survive in the SN explosion
and show distinguishing properties. In future investigations, we
will explore the properties of the companion stars after SN
explosion, which could be verified by future observations.

\section*{Acknowledgments}

We thank an anonymous referee for his/her valuable comments that
helped to improve the paper. This work is supported by the National
Natural Science Foundation of China (Grant Nos. 10873008 and
10821061), the National Basic Research Program of China (Grant Nos.
2007CB815406 and 2009CB824800), and the Yunnan Natural Science
Foundation (Grant No. 08YJ041001).

\label{lastpage}

\begin{thebibliography}{}\label{thebibliography}

\bibitem[Alexander \& Ferguson (1994)]{ale94}            Alexander D. R., Ferguson J. W., 1994, ApJ, 437, 879
\bibitem[Andronov, Pinsonneault \& Sills (2003)]{and03}  Andronov N., Pinsonneault M., Sills A., 2003, ApJ, 582, 358
\bibitem[Aubourg et al. (2008)]{aubo98}                  Aubourg E., Tojeiro R., Jimenez R., Heavens A. F., Strauss M. A., Spergel D. N., 2008, A\&A, 492, 631
\bibitem[Belczy$\acute{\rm n}$ski \& Mikolajewska (1998)]{bel98} Belczy$\acute{\rm n}$ski K., Mikolajewska J., 1998, MNRAS, 296, 77
\bibitem[Botticella et al. (2008)]{bot98}                Botticella M. T. et al.,  2008, A\&A, 479, 49
\bibitem[Cappellaro \& Turatto (1997)]{CT97}             Cappellaro E.,  Turatto M., 1997, in Ruiz-Lapuente P., Cannal R., Isern J., eds, Thermonuclear Supernovae. Kluwer, Dordrecht, P. 77
\bibitem[Chen \& Li (2007)]{che07}                       Chen W.-C., Li X.-D., 2007, ApJ, 658, L51
\bibitem[Chen \& Li (2009)]{che09}                       Chen W.-C., Li X.-D., 2009, ApJ, 702, 686
\bibitem[Chen \& Tout (2007)]{CHE07}                     Chen X., Tout C. A., 2007, Chin. J. Astro. Astrophys., 7, 245
\bibitem[Eggleton (1971)]{egg71}                         Eggleton P. P., 1971, MNRAS, 151, 351
\bibitem[Eggleton (1972)]{egg72}                         Eggleton P. P., 1972, MNRAS, 156, 361
\bibitem[Eggleton (1973)]{egg73}                         Eggleton P. P., 1973, MNRAS, 163, 279
\bibitem[Eggleton, Tout \& Fitechett (1989)]{EGG89}      Eggleton P. P., Tout C. A., Fitechett M. J., 1989, ApJ, 347, 998
\bibitem[Fedorova, Tutukov \& Yungelson (2004)]{Fed04}   Fedorova A. V., Tutukov A. V., Yungelson L. R., 2004, Astron. Lett., 30, 73
\bibitem[Goldberg \& Mazeh (1994)]{GM94}                 Goldberg D., Mazeh T., 1994, A\&A, 282, 801
\bibitem[Greggio \& Renzini (1983)]{gre83}               Greggio L., Renzini A., 1983, A\&A, 118, 217
\bibitem[Hachisu, Kato \& Luna (2007)]{hac07}            Hachisu I., Kato M., Luna G. J. M., 2007, ApJ, 659, L153
\bibitem[Hachisu, Kato \& Nomoto (1996)]{hac96}          Hachisu I., Kato M., Nomoto K., 1996, ApJ, 470, L97
\bibitem[Hachisu et al. (1999a)]{hac99a}                 Hachisu I., Kato M., Nomoto K., Umeda H., 1999a, ApJ, 519, 314
\bibitem[Hachisu, Kato \& Nomoto (1999b)]{hac99b}        Hachisu I., Kato M., Nomoto K., 1999b, ApJ, 522, 487
\bibitem[Han (1998)]{HAN98}                              Han Z., 1998, MNRAS, 296, 1019
\bibitem[Han {2008}]{han08}                              Han Z., 2008, ApJ, 677, L109
\bibitem[Han \& Podsiadlowski (2004)]{han04}             Han Z., Podsiadlowski Ph., 2004, MNRAS, 350, 1301 (HP04)
\bibitem[Han \& Podsiadlowski (2006)]{han06}             Han Z., Podsiadlowski Ph., 2006, MNRAS, 368, 1095
\bibitem[Han, Podsiadlowski \& Eggleton (1994)]{han94}   Han Z., Podsiadlowski Ph., Eggleton P. P., 1994, MNRAS, 270, 121
\bibitem[Han, Podsiadlowski \& Eggleton (1995)]{HAN95}   Han Z., Podsiadlowski Ph., Eggleton P. P., 1995, MNRAS, 272, 800
\bibitem[Han, Podsiadlowski \& Lynas-Gray (2007)]{han07} Han Z., Podsiadlowski Ph., Lynas-Gray A. E., 2007, MNRAS, 380, 1098
\bibitem[Han et al. (2003)]{HAN03}                       Han Z., Podsiadlowski Ph., Maxted P. F. L., Marsh T. R., 2003, MNRAS, 341, 669
\bibitem[Han et al. (2002)]{han02}                       Han Z., Podsiadlowski Ph., Maxted P. F. L., Marsh T. R., Ivanova N., 2002, MNRAS, 336, 449
\bibitem[Han, Tout \& Eggleton (2000)]{han00}            Han Z., Tout C. A., Eggleton P. P., 2000, MNRAS, 319, 215
\bibitem[Hansen (2003)]{hans03}                          Hansen B.~M.~S., 2003, ApJ, 582, 915
\bibitem[Hillebrandt \& Niemeyer (2000)]{hil00}          Hillebrandt W., Niemeyer J. C., 2000, ARA\&A, 38, 191
\bibitem[Hurley, Pols \& Tout (2000)]{hur00}             Hurley J. R., Pols O. R., Tout C. A., 2000, MNRAS, 315, 543
\bibitem[Hurley, Pols \& Tout (2002)]{hur02}             Hurley J. R., Pols O. R., Tout C. A., 2002, MNRAS, 329, 897
\bibitem[Iben \& Tutukov (1984)]{it84}                   Iben I. Jr., Tutukov A. V., 1984, ApJS, 54, 335
\bibitem[Iglesias \& Rogers (1996)]{igl96}               Iglesias C. A., Rogers F. J., 1996, ApJ, 464, 943
\bibitem[Justham, Wolf \& Podsiadlowski (2009)]{Jus09}   Justham S., Wolf C., Podsiadlowski P., Han Z., 2009, A\&A, 493, 1081
\bibitem[Kato \& Hachisu (2004)]{kh04}                   Kato M., Hachisu I., 2004, ApJ, 613, L129
\bibitem[King et al. (1997)]{kin97}                      King A. R., Frank J., Kolb U., Ritter H., 1997, ApJ, 484, 844
\bibitem[King, Rolfe \& Schenker (2003)]{kin03}          King A. R., Rolfe D. J., Schenker K., 2003, MNRAS, 341, L35
\bibitem[Kippenhahn \& Weigert 1967]{kip67}              Kippenhahn R., Weigert A., 1967, Z. Ap., 65, 251
\bibitem[Krishnamurthi et al. (1997)]{Kri97}             Krishnamurthi A., Pinsonneault M. H., Barnes S., Sofia S., 1997, ApJ, 480, 303
\bibitem[Kroupa, Tout \& Gilmore (1993)]{kro93}          Kroupa P., Tout C. A., Gilmore G., 1993, MNRAS, 262, 545
\bibitem[Langer et al. (2000)]{lan00}                    Langer N., Deutschmann A., Wellstein S., H\"{o}flich P., 2000, A\&A, 362, 1046
\bibitem[Lasota (2001)]{las01}                           Lasota J.-P., 2001, New Astro. Rev., 45, 449
\bibitem[Li \& van den Heuvel (1997)]{li97}              Li X.-D., van den Heuvel E. P. J., 1997, A\&A, 322, L9
\bibitem[Li \& Wang (1998)]{li98}                        Li X.-D., Wang Z.-R., 1998, ApJ, 500, 935
\bibitem[L\"{u} et al. (2009)]{lv09}                     L\"{u} G., Zhu C., Wang Z., Wang N., 2009, MNRAS, 396, 1086
\bibitem[Mannucci, Della Valle \& Panagia (2006)]{man06} Mannucci F., Della Valle M., Panagia N., 2006, MNRAS, 370, 773
\bibitem[Matteucci \& Greggio (1986)]{mat86}             Matteucci F., Greggio L., 1986, A\&A, 154, 279
\bibitem[Mazeh et al. {1992}]{MAZ92}                     Mazeh T., Goldberg D., Duquennoy A., Mayor M., 1992, ApJ, 401, 265
\bibitem[Meng, Chen \& Han (2009)]{men09}                Meng X., Chen X., Han Z., 2009, MNRAS, 395, 2103
\bibitem[Miller \& Scalo (1979)]{MS79}                   Miller G. E., Scalo J. M., 1979, ApJS, 41, 513
\bibitem[Nomoto \& Iben (1985)]{nom85}                   Nomoto K., Iben I., 1985, ApJ, 297, 531
\bibitem[Nomoto, Iwamoto \& Kishimoto (1997)]{nom97}     Nomoto K., Iwamoto K., Kishimoto N., 1997, Sci, 276, 1378
\bibitem[Nomoto, Thielemann \& Yokoi (1984)]{nom84}      Nomoto K., Thielemann F-K., Yokoi K., 1984, ApJ, 286, 644
\bibitem[Osaki (1996)]{osa07}                            Osaki Y., 1996, PASP, 108, 39
\bibitem[Paczy\'{n}ski (1976)]{PAC76}                    Paczy\'{n}ski B., 1976, in Eggleton P. P., Mitton S., Whelan J., eds. Structure and Evolution of Close Binaries. Kluwer, Dordrecht, p. 75
\bibitem[Patat et al. (2007)]{pat07}                     Patat F. et al., 2007, Sci, 317, 924
\bibitem[Perlmutter et al. (1999)]{per99}                Perlmutter S. et al., 1999, ApJ, 517, 565
\bibitem[Podsiadlowski et al. (2008)]{pod08}             Podsiadlowski Ph., Mazzali P., Lesaffre P., Han Z., F\"{o}rster F., 2008, New Astro. Rev., 52, 381
\bibitem[Podsiadlowski, Rappaport \& Pfahl (2002)]{PO02} Podsiadlowski Ph., Rappaport S., Pfahl E. D., 2002, ApJ, 565, 1107
\bibitem[Pols et al. (1998)]{pol98}                      Pols O. R., Schr\"{o}der K. P., Hurly J. R., Tout C. A., Eggleton P. P., 1998, MNRAS, 298, 525
\bibitem[Pols et al. (1995)]{pol95}                      Pols O. R., Tout C. A., Eggleton P. P., Han Z., 1995, MNRAS, 274, 964
\bibitem[Pols et al. (1997)]{POL97}                      Pols O. R., Tout C. A., Schr\"{o}der K. P., Eggleton P. P., Manners J., 1997, MNRAS, 289, 869
\bibitem[Riess et al. (1998)]{rie98}                     Riess A. et al., 1998, AJ, 116, 1009
\bibitem[R\"{o}pke \& Hillebrandt (2005)]{rop05}         R\"{o}pke F. K., Hillebrandt W., 2005, A\&A, 431, 635
\bibitem[Ruiter, Belczynski \& Fryer (2009)]{Ruit04}     Ruiter A. J., Belczynski K., Fryer C. L., 2009, ApJ, 699, 2026
\bibitem[Ruiz-Lapuente et al.(2004)]{Ruiz04}             Ruiz-Lapuente P. et al., 2004, Nat, 431, 1069
\bibitem[Saio \& Nomoto (1985)]{sai85}                   Saio H., Nomoto K., 1985, A\&A, 150, L21
\bibitem[Schawinski (2009)]{Scha09}                      Schawinski K., 2009, MNRAS, 397, 717
\bibitem[Schr\"{o}der, Pols \& Eggleton (1997)]{Schr97}  Schr\"{o}der K. P., Pols O. R., Eggleton P. P., 1997, MNRAS, 285, 696
\bibitem[Shatsky \& Tokovinin (2002)]{Sha02}             Shatsky N., Tokovinin A., 2002, A\&A, 382, 92
\bibitem[Sills, Pinsonneault \& Terndrup (2000)]{Sil00}  Sills A., Pinsonneault M. H., Terndrup D. M., 2000, ApJ, 534, 335
\bibitem[Smak (1983)]{sma83}                             Smak J., 1983, ApJ, 272, 234
\bibitem[Sokoloski et al. (2006)]{sok06}                 Sokoloski J. L., Luna G. J. M., Mukai K., Kenyon S. J., 2006, Nat, 442, 276
\bibitem[Timmes, Diehl \& Hartmann (1997)]{tim97}        Timmes F. X., Diehl R., Hartmann D. H., 1997, ApJ, 479, 760
\bibitem[Timmes, Woosley \& Taam (1994)]{tim94}          Timmes F. X., Woosley S. E., Taam R. E., 1994, ApJ, 420, 348
\bibitem[Totani et al. (2008)]{tim94}                    Totani T., Morokuma T., Oda T., Doi M., Yasuda N., 2008, PASJ, 60, 1327
\bibitem[Uenishi, Nomoto \& Hachisu (2003)]{uen03}       Uenishi T., Nomoto K., Hachisu I., 2003, ApJ, 595, 1094
\bibitem[Umeda et al. (1999)]{ume99}                     Umeda H., Nomoto K., Yamaoka H., Wanajo S., 1999, ApJ, 513, 861
\bibitem[van den Bergh \& Tammann (1991)]{vand91}        van den Bergh S., Tammann G. A., 1991, ARA\&A, 29, 363
\bibitem[van Paradijs (1996)]{vanp96}                    van Paradijs J., 1996, ApJ, 464, L139
\bibitem[Voss \& Nelemans (2008)]{Vos08}                 Voss R., Nelemans G., 2008, Nat, 451, 802
\bibitem[Wang et al. (2009a)]{wan09a}                    Wang B., Meng X., Chen X., Han Z., 2009a, MNRAS, 395, 847
\bibitem[Wang et al. (2009b)]{wan09b}                    Wang B., Chen X., Meng X., Han Z., 2009b, ApJ, 701, 1540 (WCMH09)
\bibitem[Wang et al. (2008a)]{wan08a}                    Wang B., Meng X., Wang X.-F., Han Z., 2008a, Chin. J. Astro. Astrophys., 8, 71
\bibitem[Wang et al. (2008b)]{wan08b}                    Wang X.-F. et al., 2008b, ApJ, 675, 626
\bibitem[Warner (1995)]{war95b}                          Warner B., 1995, Cataclysmic Variable Stars. Cambridge, England: Cambridge University Press
\bibitem[Webbink (1984)]{web84}                          Webbink R. F., 1984, ApJ, 277, 355
\bibitem[Whelan \& Iben (1973)]{whe73}                   Whelan J., Iben I., 1973, ApJ, 186, 1007
\bibitem[Willems \& Kolb (2004)]{WK04}                   Willems B., Kolb U., 2004, A\&A, 419, 1057
\bibitem[Xu \& Li (2009)]{xu09}                          Xu X.-J., Li X.-D., 2009, A\&A, 495, 243
\bibitem[Yoon \& Langer (2005)]{yoo05}                   Yoon S.-C., Langer N., 2005, A\&A, 435, 967
\bibitem[Yoon, Langer \& Scheithauer(2004)]{y04}         Yoon S.-C., Langer N., Scheithauer S., 2004, A\&A, 425, 217
\bibitem[Yungelson \& Livio (1998)]{YUN98}               Yungelson L., Livio M., 1998, ApJ, 497, 168
\bibitem[Zoccali et al. (2000)]{zoc00}                   Zoccali M., Cassisi S., Frogel J. A., Gould A., Ortolani S., Renzini A., Rich R. M., Stephens A. W., 2000, ApJ, 530, 418

\end{thebibliography}
\end{document}